\documentclass[aps,prl,twocolumn,showpacs,groupedaddress,floatfix]{revtex4}

\usepackage{graphicx}
\usepackage{dcolumn}
\usepackage{bm}
\usepackage{amssymb}

\begin{document}

\newcommand{\dzero}     {D0}
\newcommand{\met}       {\mbox{$\not\!\!E_T$}}
\newcommand{\deta}      {\mbox{$\eta^{\rm det}$}}
\newcommand{\meta}      {\mbox{$\left|\eta\right|$}}
\newcommand{\mdeta}     {\mbox{$\left|\eta^{\rm det}\right|$}}
\newcommand{\rar}       {\rightarrow}
\newcommand{\rargap}    {\mbox{ $\rightarrow$ }}
\newcommand{\tbbar}     {\mbox{$t\bar{b}$}}
\newcommand{\tqbbar}    {\mbox{$tqb$}}
\newcommand{\ttbar}     {\mbox{$t\bar{t}$}}
\newcommand{\bbbar}     {\mbox{$b\bar{b}$}}
\newcommand{\ccbar}     {\mbox{$c\bar{c}$}}
\newcommand{\qqbar}     {\mbox{$q\bar{q}$}}
\newcommand{\ppbar}     {\mbox{$p\bar{p}$}}
\newcommand{\comphep}   {\sc{c}\rm{omp}\sc{hep}}
\newcommand{\herwig}    {\sc{herwig}}
\newcommand{\pythia}    {\sc{pythia}}
\newcommand{\alpgen}    {\sc{alpgen}}
\newcommand{\singletop} {\sc{SingleTop}}
\newcommand{\reco}      {\sc{reco}}
\newcommand{\Mchiggs}   {\mbox{$M({\rm jet1,jet2},W)$}}
\newcommand{\geant}     {\sc{geant}}
\newcommand{\gt}     	{>}
\newcommand{\lt}     	{<}
\newcommand{\eq}     	{=}

\hspace{5.2in}\mbox{Fermilab-Pub-08-229-E}

\title{Search for charged Higgs bosons decaying to top and bottom quarks in $\mathbf{p}\bar{\mathbf{p}}$ collisions}

%
\author{V.M.~Abazov$^{36}$}
\author{B.~Abbott$^{75}$}
\author{M.~Abolins$^{65}$}
\author{B.S.~Acharya$^{29}$}
\author{M.~Adams$^{51}$}
\author{T.~Adams$^{49}$}
\author{E.~Aguilo$^{6}$}
\author{M.~Ahsan$^{59}$}
\author{G.D.~Alexeev$^{36}$}
\author{G.~Alkhazov$^{40}$}
% ...

\affiliation{\vspace{0.1 in}(The D\O\ Collaboration)\vspace{0.1 in}}
% ...
\affiliation{$^{6}$University of Alberta, Edmonton, Alberta, Canada,
                Simon Fraser University, Burnaby, British Columbia, Canada,
                York University, Toronto, Ontario, Canada, and
                McGill University, Montreal, Quebec, Canada}
% ...
\affiliation{$^{29}$Tata Institute of Fundamental Research, Mumbai, India}
% ...
\affiliation{$^{36}$Joint Institute for Nuclear Research, Dubna, Russia}
% ...
\affiliation{$^{40}$Petersburg Nuclear Physics Institute,
                St. Petersburg, Russia}
% ...
\affiliation{$^{49}$Florida State University, Tallahassee, Florida 32306, USA}
% ...
\affiliation{$^{51}$University of Illinois at Chicago,
                Chicago, Illinois 60607, USA}
% ...
\affiliation{$^{59}$Kansas State University, Manhattan, Kansas 66506, USA}
% ...
\affiliation{$^{65}$Michigan State University,
                East Lansing, Michigan 48824, USA}
% ...
\affiliation{$^{75}$University of Oklahoma, Norman, Oklahoma 73019, USA}
% ...

\date{July 5, 2008}

\begin{abstract}

We describe a search for production of a charged Higgs
boson, $q \bar{q}^{\prime} \rar H^{+}$, reconstructed in the 
$t\bar{b}$ final state in the mass range $180 \le M_{H^{+}} \le 300$~GeV. 
The search was undertaken at the Fermilab Tevatron collider with a 
center-of-mass energy $\sqrt{s}~\eq~1.96$~TeV and uses 0.9~fb$^{-1}$ of 
data collected with the D0 detector.  
We find no evidence for charged Higgs boson production and set upper
limits on the production cross section in the Types~I, II and III
two-Higgs-doublet models (2HDMs). An excluded region in the
$(M_{H^+},\tan\beta)$ plane for Type~I 2HDM is presented.

\end{abstract}

\maketitle

In the standard model (SM), one $SU(2)$ doublet induces electroweak symmetry 
breaking, which leads to a single elementary scalar particle: the neutral 
Higgs boson. Two $SU(2)$ doublets perform the task of electroweak symmetry 
breaking in two-Higgs-doublet models (2HDMs)~\cite{hhg:1990}. This
leads to five physical Higgs bosons among 
which two carry charge.  Hence the discovery of a charged Higgs boson
would be unambiguous evidence of new physics beyond the SM. Various types 
of 2HDMs are distinguished by their strategy for avoiding flavor-changing 
neutral currents (FCNCs). In the Type~I 2HDM, only one of these doublets couples 
to fermions. In the Type~II 2HDM, a symmetry is imposed so that one doublet 
couples to up-type fermions and the other couples to down-type
fermions; an approach used in minimal supersymmetry
extensions~\cite{hhg:1990}. In
Type~III 2HDMs, both doublets couple to fermions, no symmetry is imposed and 
FCNCs are avoided by other methods. For example, in one Type~III model, FCNCs
are suppressed by the small mass of the first and second generation
quarks~\cite{hy:1999}. 

In this Letter we present the first search for a charged Higgs 
boson ($H^+$) directly produced by quark-antiquark annihilation, 
and decaying into the $t\bar{b}$~\cite{bib:notation} final state,
in the $180 \le M_{H^+} \le 300$~GeV mass range.
In most models this decay dominates for large regions of
parameter space when the $H^+$ mass ($M_{H^+}$) is greater than 
the mass of the top quark ($m_{t}$). 
Exploring the mass range $M_{H^+} \gt m_{t}$ is complementary to 
previous Tevatron searches~\cite{TevatronLim} that have been 
performed in top quark decays for the $M_{H^+} \lt m_{t}$ 
region.
We analyze 0.9~fb$^{-1}$ of data from \ppbar~collisions
at a center-of-mass energy of $\sqrt{s}~\eq~1.96$~TeV recorded
from August 2002 to December 2006 using the \dzero\ detector~\cite{UpgradedD0:2006}.
Since the {\dzero} single top quark analysis~\cite{abazov:181802}
reconstructs precisely the same final state in the $s$-channel
$W^+ \rar t\bar{b}$ process, we use the 
dataset from
that search.

Direct searches for a charged Higgs boson have been performed at the
CERN $e^+e^-$ collider (LEP)~\cite{LEP:2001} and the
Fermilab Tevatron collider~\cite{TevatronLim}, while indirect 
searches have been undertaken at the $B$
factories~\cite{BfactLim:2007,bib:Btaunu}. No evidence for $H^+$
has been found so far.
Limits on the charged Higgs mass and the
ratio of vacuum expectation values of the two Higgs fields
($\tan\beta$) are typically calculated in the context of the Type~II
2HDM~\cite{pdg:2007}. 
The combined results from the LEP  
experiments and those from $B$ factories yield
$M_{H^+} \gt 78.6$~GeV~\cite{pdg:2007} and 
$M_{H^+} \gt 295$~GeV~\cite{BfactLim:2007}, respectively, at the 95\%
C.L. and assuming Type~II 2HDM.

The charged Higgs Yukawa couplings carry information about new physics
beyond the SM and it 
has been noted that 2HDM couplings in Types~I and II 2HDM
can be quite large~\cite{roy:2004}. For a Type~III 2HDM, large
contributions from heavy quark-antiquark annihilation can 
be expected if the top-quark/charm-quark mixing parameter ($\xi^{U}_{tc}$) is large
\cite{hy:1999}.  
In many models, if $M_{H^+} \gt m_{t}$, then the branching fraction of the
charged Higgs boson 
to $t\bar{b}$ is of order unity, owing to the mass dependence
of the couplings and the large top quark mass. 

We use the program {\comphep}~\cite{comphep} to
simulate charged Higgs boson production and selected decay  
$q\bar{q}^\prime \rar H^+ \rar t\bar{b} \rar W^+ b
\bar{b} \rar \ell^+\nu b\bar{b}$ where $\ell$ represents an electron or
muon. This is
done for seven $M_{H^+}$ values ranging from $180$ to  
$300$~GeV.  The lower mass value is dictated by the kinematics of the
decay $H^+ \rar t\bar{b}$ which requires $M_{H^+} \gt m_{t}+m_{b}$, 
where $m_{b}$ is the mass of the bottom quark.  
The upper mass value is chosen based on the fact that, in this mass
range,  the production
cross section decreases by approximately an order of 
magnitude for any of the models considered.
The couplings are set to produce pure chiral state samples
that are combined in different proportions to simulate the desired 2HDM type. 
The size of the interference term proportional to the product of
the left and right-handed couplings is considered negligible.
The size of this interference term 
is of order 1\% of the total amplitude in the $\tan\beta \lt 30$ region for the
Type~II 2HDM, much less than 1\% for the Type~I 2HDM and
non-relevant for a Type~III 2HDM.
Each choice of couplings 
determines the total width, $\Gamma_{H^+}$, and the initial-state 
quark flavor composition.  This quark flavor
composition of the signal samples is determined by the
value of the element $|V_{ij}|$ of the Cabibbo-Kobayashi-Maskawa (CKM)
matrix~\cite{ckm}  
and the CTEQ6L1 parton distribution functions (PDFs)~\cite{cteq}.
In these simulated signal samples, $\Gamma_{H^+}$ ranges from approximately
$4$~GeV for $M_{H^+} \eq 180$~GeV to $9$~GeV for $M_{H^+} \eq 300$~GeV.

In order to simulate the kinematic distributions of a particular
model, the left-handed and right-handed signal samples are combined
with event weights equal to the fraction of the production cross
section associated with the left-handed or right-handed coupling contribution.
The Type~II 2HDM couplings for right-handed ($R$) and left-handed ($L$) chiral 
states are $V_{CKM}^{qq'} g m_{q'}\tan\beta / (\sqrt{2} M_W)$ and $V_{CKM}^{qq'}g
m_{q}\cot\beta / (\sqrt{2} M_W)$, 
where $V_{CKM}^{qq'}$ is the CKM matrix element, $m_{q}/m_{q'}$ the up/down-type
quark mass, $M_W$ the mass of the $W$ boson and $g$ the SM weak coupling
constant. The $R$($L$) couplings 
in Type~I and III 2HDMs are 
$V_{qq'}g m_{q'}\tan\beta / (\sqrt{2} M_W)$ ($-V_{qq'}g m_{q}\tan\beta
/ (\sqrt{2}M_W)$) and $-(V_{\rm CKM}\hat{Y}_D)_{qq'}$
($(\hat{Y}_U^\dagger V_{\rm CKM})_{qq'}$), where  
$\hat{Y}_{ij}^{U,D}{\eq}\xi\sqrt{2 m_i m_j}/v$, $v$ is the vaccum
expectation value and $\xi$ is taken as a
free parameter of the model.
For the simulation of Type~I 2HDM, left-handed and right-handed samples are added in
equal proportion.  For the simulation of Type~II 2HDM, signal samples are combined
to simulate four $\tan\beta$ values or ranges: $\tan\beta \lt 0.1$,
$\tan\beta \eq 1$,   $\tan\beta \eq 5$, and $\tan\beta \gt 10$. The Type~I 2HDM
and $\tan \beta \eq 1$ Type~II models share the same left/right-handed proportions. 
For the Type~III 2HDM as described in \cite{hy:1999}, quark-antiquark annihilation
is dominated by right-handed couplings. This
model is simulated using the same proportions of left-handed and right-handed
samples as used to simulate the $\tan\beta \gt 10$ Type~II model.
This approach provides an adequate simulation of signal event
kinematics
only for model parameter values that result in a charged Higgs width
comparable or smaller than the experimental mass resolution of ${\cal O}(10)$~GeV.

Background contributions from $W+$jets and top quark pair (\ttbar) production are
modeled using the {\alpgen} Monte Carlo (MC) event generator~\cite{alpgen}.
The single top quark samples are generated with the {\singletop}~\cite{singletop}
MC event generator.
For both samples, we assume a top
quark mass of $175$~GeV and use the CTEQ6L1 PDFs.
After generation, the events are passed through a {\geant}-based
simulation~\cite{geant}  
of the \dzero\ detector and subsequently through standard
reconstruction procedures that  
correct differences between the simulation and data.

The background contribution from misreconstructed multijet events is 
modeled using data events containing misidentified leptons and is normalized
to the signal data together with the $W+$jets sample, which contains leptons from the $W$ boson decay~\cite{abazov:181802}.

We search for charged Higgs bosons in the 
$H^+ \rar t\bar{b} \rar \ell^{+} \nu b \bar{b}$ final state, and
hence require that events satisfy triggers with a jet and an
electron or muon.
Selections that are identical to the two-jet analysis channel
for the {\dzero} single top quark analysis~\cite{abazov:181802}
are imposed on each observable in the data, background and 
charged Higgs boson signal samples to select events with 
$t\bar{b}$ final state signatures.  Events are required to  have
a primary vertex with three or more tracks attached and a lepton originating from the 
primary vertex~\cite{abazov:181802}.
The electron (muon) channel selection requires only one isolated electron (muon) with 
$E_{T} \gt 15$ ($p_{T} \gt 18$)~GeV within the pseudorapidity region
$|\eta| \lt 1.1$ (2.0).
Events with two isolated leptons are rejected.
For both channels, events are required to have missing
transverse energy within $15 \lt $~\mbox{$\not\!\!E_T$}~$\lt 200$~GeV.
We require that events have exactly two jets, with the highest $p_{T}$
jet satisfying $p_{T} \gt 25$~GeV and $|\eta| \lt 2.5$, and
the second jet satisfying $p_{T} \gt 20$~GeV and $|\eta| \lt 3.4$.

Since both jets of the signal events are $b$~jets, we select data
events having one or two jets identified as such via a neural 
network-based tagging algorithm~\cite{TScanlon}.
MC simulated events are weighted using a $b$-tag probability
derived from data.
The signal acceptances after the complete selection increase 
monotonically in the mass range $200 \lt M_{H^+} \lt 300$~GeV, for example, 
from $(0.48 \pm 0.06)$\% to $(1.24 \pm 0.20)$\% for
$\tan\beta<0.1$, statistical and systematic uncertainties included.
The signal acceptances for a 
given $M_{H^+}$ decreases by at most 0.12\% with increasing $\tan\beta$.

\begin{figure}
\includegraphics[scale=0.35]{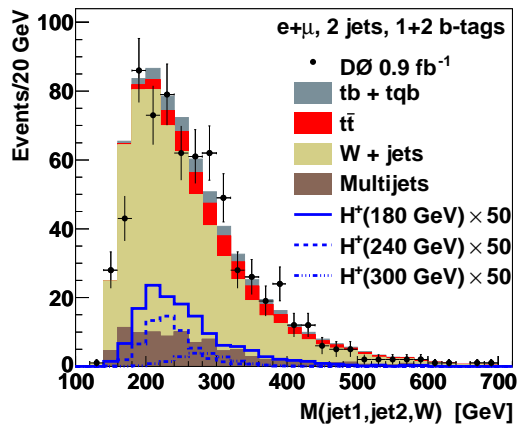}
\caption{\label{fig:plots} Distribution of the discriminating variable, \Mchiggs, for
  the signal, background model and data, for the combined electron and muon channels
  with exactly two jets and with one or two $b$ tags.
  The signal distributions correspond to
  a Type~III 2HDM for charged Higgs boson masses
  $180,240,300$~GeV, and are normalized according to the
  production cross section presented in Ref.~\cite{hy:1999} scaled by
  a factor of 50.  
} 
\end{figure}

A distinctive feature of signal events is the large mass of the
charged Higgs boson.  We therefore use the reconstructed invariant
mass of the top and bottom quark system as the discriminating 
variable for the charged Higgs signal. 
We define this variable as the invariant mass \Mchiggs.
In the reconstruction of the $W$ boson, there are up to two possible solutions for
the neutrino momentum component along the beam axis ($p_z$). In these cases, the solution 
with the smallest absolute value of the $p_z$ momentum is chosen.
Figure~\ref{fig:plots} shows the \Mchiggs~distribution   
after selection, with an example signal normalized 
to the production cross section for a Type~III 2HDM~\cite{hy:1999}
and for three different mass values. 

The data yield for all analysis channels combined amounts to 697 
events, after the complete selection. Similarly, for the sum of 
all background sources, the total expected yield is 721$\pm$42. 
For the separate background sources, the yields are 531 for 
$W+$jets, 95 for multijets, 59 for \ttbar ~and 36 for the single 
top background.

The systematic uncertainties on the signal and background model are
estimated using the methods described in Ref.~\cite{abazov:181802}.
Two of the dominant sources of systematic uncertainty arise from the 
jet energy scale (JES) correction uncertainty and the uncertainty on 
the $b$-tag rates applied to MC events (described above). 
For the $H^+$ signal, the uncertainty on the model-dependent proportion of initial-state 
parton flavor contribution plays a dominant role. Simulated signal
events with different exclusive initial-state quark combinations are 
used to assess the latter source of uncertainty. A value of 10\% is assigned based on
variations in yield and shape of the reconstructed invariant mass
distribution. 


\begin{table}[t!!!]
\caption{\label{table:obslim} Observed limits on the production cross
  section (in pb) times branching fraction 
  $\sigma(q \bar{q}^\prime \rar H^+) \times \mathcal{B}(H^+ \rar t
  \bar{b})$. The expected limits are shown in  
  parenthesis for comparison. These limits apply to the Type~II 2HDM. 
  The limits obtained for $\tan\beta \eq 1$ and
  $\tan\beta \gt 10$ are also valid for Type~I and  
  Type~III 2HDMs, respectively.  Limits shown in square brackets are
  only valid for the general production of a charged scalar via a
  purely left-handed coupling with width smaller
  than the experimental resolution.  These limits are not valid for the
  production of a charged Higgs boson in Type~II 2HDM since the
  charged Higgs width is expected to be larger than the experimental resolution.
}
\begin{ruledtabular}
\begin{tabular}{c|cccc}
$M_{H^+}$~(GeV) 	&  {$\rm tan\beta<0.1$}	& {$\rm tan\beta=1$}	& {$\rm tan\beta=5$} 	& {$\rm tan\beta>10$} \\
\hline
~~$180$ &   
12.9 (11.4) &
14.3 (12.2) &
13.7 (11.7) &
13.7 (12.2) \\
~~$200$ &   
[ 5.9 (9.6) ] & 
6.3 (9.9) &
6.5 (10.0) &
6.5 (10.0) \\
~~$220$ &   
[ 2.9 (4.2) ] &
3.0 (4.4) &
3.0 (4.5) &
3.0 (4.5) \\
~~$240$ &   
[ 2.3 (3.1) ] &
2.4 (3.3) &
2.6 (3.5) &
2.6 (3.5) \\
~~$260$ &   
[ 3.0 (2.8) ]&
3.0 (2.9) &
3.0 (3.0) & 
3.0 (3.0) \\
~~$280$ &   
[ 4.0 (2.6) ] &
4.2 (2.7) &
4.5 (2.9) &
4.5 (2.9) \\
~~$300$ &   
[ 4.5 (2.4) ]&
4.7 (2.4) &
4.9 (2.5) &
4.9 (2.5) \\
\end{tabular}
\end{ruledtabular}
\end{table}

We observe no excess of data over background and proceed to set 
upper limits on $H^+$ boson production.  We construct a
binned likelihood function and use Bayesian statistics to calculate upper 
limits on the signal production cross section times the branching
fraction ($\sigma\times \mathcal{B}$) to the $t\bar{b}$ final state. 
A flat positive prior is
used for the signal cross section.  All sources of systematic
uncertainty and their correlations are taken into account in calculating
$\sigma\times \mathcal{B}$ upper limits for different 2HDM types
at the 95\%~C.L. At the level of precision reported, the observed limits are insensitive to changes in top mass in the range $170 < m_{t} < 175$ GeV. The observed and expected $\sigma\times \mathcal{B}$
limits are reported in Table~\ref{table:obslim}.

\begin{figure}[t]
\includegraphics[scale=0.35]{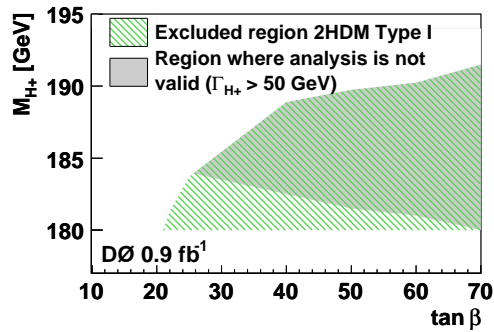}
\caption{\label{fig:results} The 95\%~C.L. excluded region in the $M_{H^+}$ vs $\tan\beta$
  space for Type~I 2HDM. The region for which $\Gamma_{H^+} \gt 50$~GeV
  indicates the approximate area where the charged Higgs width is significantly
  larger than the detector resolution and hence the analysis
  is not valid.}
\end{figure}

The $\sigma \times \mathcal{B}$ upper limits obtained are compared to
the expected signal cross section in the Type~I 2HDM to exclude a
region of the $M_{\rm H^+}$ and $\tan\beta$ parameter space, shown in
Fig.~\ref{fig:results}.  The analysis 
sensitivity is currently not sufficient to exclude regions of
$\tan\beta \lt 100$ in the Type~II 2HDM.  In a Type~III
2HDM~\cite{hy:1999}, the charged Higgs boson width depends
quadratically on the mixing parameter $\xi$.
This limits our ability
to exclude regions in the $M_{\rm H^+}$ and $\xi$ parameter space.

In summary, we have performed the first direct search for the production of
charged Higgs bosons in the reaction $q\bar{q}' \rar H^+ \rar t\bar{b}$ and 
we have presented limits on the production cross section times
branching fraction for Types~I, II and III 2HDMs in the mass range
$180 \le M_{H^+} \le 300$~GeV.  A region in 
the $M_{H^+}$ vs $\tan\beta$ plane has been excluded at the 95\%~C.L. for 
Type~I 2HDMs.

%
We thank the staffs at Fermilab and collaborating institutions, 
and acknowledge support from the 
DOE and NSF (USA);
CEA and CNRS/IN2P3 (France);
FASI, Rosatom and RFBR (Russia);
CNPq, FAPERJ, FAPESP and FUNDUNESP (Brazil);
DAE and DST (India);
Colciencias (Colombia);
CONACyT (Mexico);
KRF and KOSEF (Korea);
CONICET and UBACyT (Argentina);
FOM (The Netherlands);
STFC (United Kingdom);
MSMT and GACR (Czech Republic);
CRC Program, CFI, NSERC and WestGrid Project (Canada);
BMBF and DFG (Germany);
SFI (Ireland);
The Swedish Research Council (Sweden);
CAS and CNSF (China);
and the
Alexander von Humboldt Foundation (Germany).
%



\begin{thebibliography}{99}
\bibitem{bib:notation}We use the $H^+$ notation to refer to both $H^+$
  and its charge conjugate state $H^-$.  Similarly, the $t\bar{b}$ notation is used here to represent both the $t\bar{b}$ state and its charge conjugate state $\bar{t}b$.
\end{thebibliography}
\end{document}